\documentstyle{article}
\title{Crystalline Spinon Basis for RSOS Models}
\author{Atsushi Nakayashiki\
and Yasuhiko Yamada\\
Graduate School of Mathematics\\
Kyushu University}
\date{}

\begin{document}
\def\uq{U_q(\widehat{sl_2})}
\def\be{{\bf \epsilon}}
\def\la{\lambda}
\def\La{\Lambda}
\def\c{\varphi^\ast}
\def\v{\varphi}
\def\ot{\otimes}
\def\bz{{\bf Z}}
\def\et{\tilde{e}}
\def\ft{\tilde{f}}
\def\jt{\tilde{j}}
\def\ep{\epsilon}
\def\ca{{\cal A}}
\def\pek{{\cal P}^k}
\def\pel{{\cal P}^l}
\def\peN{{\cal P}^N}
\def\enpn{\epsilon(j_{n+1}+j_n)}
\def\ennm{\epsilon(j_n+j_{n-1})}
\def\knpn{K(p_{n+1},p_n)}
\def\knnm{K(p_n,p_{n-1})}
\def\cb{{\cal B}}
\def\br{r}
\def\th{\tilde{H}}
\def\cp{{\cal P}}
\def\bp{\bar{p}}
\def\ra{\rightarrow}
\def\teta{\tilde{\eta}}
\def\txi{\tilde{\xi}}
\newtheorem{lemma}{Lemma}
\newtheorem{sublem}{Sublemma}
\newtheorem{cor}{Corollary}
\newtheorem{definition}{Definition}
\newtheorem{theorem}{Theorem}
\newtheorem{prop}{Proposition}
\newtheorem{conj}{Conjecture}
\maketitle
\begin{abstract}
The crystalline spinon basis for the RSOS models associated with
$\widehat{sl_2}$ is studied. This basis gives fermionic type character
formulas for the branching coefficients of the coset
$(\widehat{sl_2})_l \times (\widehat{sl_2})_N/(\widehat{sl_2})_{l+N}$.
In addition the path description
of the parafermion characters is found as a limit of the spinon description
of the string functions.
\end{abstract}
\par

\vskip4mm
\noindent
{\Large\bf 0 \hskip4mm Introduction}
\vskip4mm
\noindent
The notion of crystalline spinons was introduced and studied
in \cite{NY} for the case of the higher spin XXZ model.
Employing this notion yields a new parametrization
of a base of the integrable highest weight $\uq$ modules which
naturally leads to the fermionic type character formulas for these
modules proposed in \cite{BLS2}.
In this paper we study the crystalline spinon basis
for RSOS models associated with $\widehat{sl_2}$.
The idea investigated here is similar
to that in \cite{NY}. Let us briefly explain this idea in a way
that allows comparison to the case of the XXZ model.

The space of states of the RSOS quantum spin chain is
\begin{eqnarray}
&&
{\cal W}=\oplus_p {\bf C} \ p,
\nonumber
\end{eqnarray}
where the sum is over all level $k$ infinite restricted paths
$p=[\cdots p_1,p_0,p_{-1},\cdots]$ satisfying some boundary conditions.
A representation theoretical description of this space
in the regime III\cite{ABF,DJKMO}
similar to the description of anti-ferromagnetic XXZ models \cite{DFJMN,IIJMNT}
has been given \cite{JMO,DJO}. We have:
\begin{eqnarray}
&&
{\cal H}=\oplus\{\hbox{ $\uq$ singlet}\}
\subset
\oplus V(\xi)\ot V(\eta)\ot V(\eta')^{\ast a}\ot V(\xi')^{\ast a},
\nonumber
\end{eqnarray}
where $\xi,\xi'$ and $\eta,\eta'$ are level $l$ and $N$ $(l+N=k)$
dominant integral weights, respectively.
The creation and annihilation operators are given in terms
of vertex operators, and their commutation relations are determined \cite{JMO}.
Together, these suggest the following particle picture of the space of states
\cite{JMO,BR}
\begin{eqnarray}
&&
{\cal F}=
\oplus_{n=0}^\infty
\Big[
\oplus_{p,p'} \
{\bf C}((z_1,\cdots,z_n))\ot
[p_n,\cdots,p_1]\ot[p_n^\prime,\cdots,p_1^\prime]
\Big]^{\hbox{sym}},
\nonumber
\end{eqnarray}
where $p$ and $p'$ run over all restricted paths of level
$l$ and $N$, respectively, and sym represents symmetrization
w.r.t. the S-matrix.

Our main aim in this paper is to establish rigorously the
equality ${\cal H}={\cal F}$ at $q=0$.
By considering the set of elements of the form
$b\ot b^\prime\ot b_{-\eta^\prime}\ot b_{-\xi^\prime}$
in ${\cal H}\vert_{q=0}$, we obtain a crystalline
spinon description of the highest weight elements
in the tensor products of two integrable highest weight $\uq$
modules, where $b_{-\eta^\prime}$ and $b_{-\xi^\prime}$ are
the lowest weight elements with lowest weights $-\eta^\prime$
and $-\xi^\prime$ respectively.
As a corollary of
this parametrization, we obtain the fermionic type formula
for the branching coefficients.

In the course of examining the character formulas obtained
from spinon descriptions in \cite{NY} and in this paper,
we have found a path description of the character of the parafermion
space. Interestingly, this path description is nothing but
the one dimensional configuration sum of the ABF model in regime II
in the thermodynamic limit. The evaluation of this sum is the most complicated
computation in \cite{ABF}.

The present paper is organized in the following manner.
In section 1 we review the RSOS models in the framework
of \cite{JMO,DJO}. The crystalline
spinon basis for XXZ models is reviewed in section 2.
In section 3 we introduce the crystalline creation algebra for the RSOS
models.
The crystalline spinon description for the space of sates of the RSOS
models is given in section 4.
In section 5 we explain how the spinon formulas of the string functions
naturally lead to a path description of the parafermion characters.
We briefly explain the derivation of the crystalline creation algebra
in Appendix A.

\section{Review of the RSOS model in the representation theoretical
formulation}
\par
Here we review the formulation of the RSOS model given by
Jimbo-Miwa-Ohta \cite{JMO} and introduce notation.
We adopt the notation of \cite{JMO} unless otherwise stated.
Let us fix an integer $k$ satisfying $1\leq N\leq k-1$ and set $l=N-k$.
The spin variable of the RSOS model takes the level $k$ dominant
integral weights.
The Boltzmann weight is given by
$$
\begin{array}{ccc}
\la & - & \mu \\
\vert & & \vert \\
\mu' & - & \nu
\end{array}
=
W^N_k( \begin{array}{cc} \la & \mu \\ \mu' & \nu \end{array}
\vert z),
$$
where
$W^N_k(
\begin{array}{cc} \la & \mu \\ \mu' & \nu \end{array}
\vert z)$ is determined from the commutation relations
of vertex operators,
$$
\check{\bar{R}}_{NN}({z_1 \over z_2})
\Phi^{\nu V^{(N)}_1}_{\mu}(z_1)
\Phi^{\mu V^{(N)}_2}_{\la}(z_2) =
\sum_{\mu'}\Phi^{\nu V^{(N)}_2}_{\mu}(z_2)
\Phi^{\mu V^{(N)}_1}_{\la}(z_1)
W^N_k(
\begin{array}{cc} \la & \mu \\ \mu' & \nu \end{array}
\vert {z_1\over z_2}).
$$
Here the Boltzmann weight is zero unless the pairs
$(\la,\mu),(\mu,\nu),(\nu,\mu'),(\mu',\la)$ are admissible.
The admissibility condition is specified by the existence
condition of the vertex operators.
Let us denote by $\la_j^{(k)}=(k-j)\La_0+j\La_1$ the level
$k$ dominant integral weight and set
$P^0_k=\{\la_j^{(k)} \vert 0\leq j\leq k\}$.

\begin{definition}
The pair $(\la_j^{(k)},\la_{j'}^{(k)})$
is called admissible if the following
conditions are satisfied
\begin{eqnarray}
&&
j-j'\in\{N,N-2,\cdots, -N\},
\quad
N\leq j+j'\leq 2k-N.
\nonumber
\end{eqnarray}
\end{definition}
\vskip3truemm

\noindent
Let us define the bijection $\sigma$ of $P^0_N$ by
$B(\eta)\ot B^{(N)}\simeq B(\sigma(\eta))$.
Explicitly, if $\eta=\la^{(N)}_j$
then $\sigma(\eta)=\la^{(N)}_{N-j}$.
The admissible pairs are parametrized as described below.

\begin{prop}
There is a bijection
$$
P^0_{l}\times P^0_{N}\simeq\{\hbox{the admissible pairs}\}
$$
given by
$$
(\xi,\eta)\mapsto (\xi+\eta,\xi+\sigma(\eta)).
$$
\end{prop}
\vskip3truemm

\noindent
Take any $(\xi,\eta),(\tilde{\xi},\tilde{\eta})\in P^0_{l}\times P^0_{N}$.
Then we state

\begin{definition}
$a=(a(n))_{n\in\bz}$ is called a
$(\xi\eta,\tilde{\xi}\tilde{\eta})$ restricted path if
\begin{description}
\item[(i)] $a(n)\in P^0_k$, and $(a(n),a(n+1))$ is admissible
for any $n$.
\item[(ii)]
$
a(n)=\xi+\sigma^n(\eta)\quad(n>>0),
\hbox{ and }
a(n)=\txi+\sigma^n(\teta)\quad(n<<0).
$
\end{description}
\end{definition}
\vskip3truemm

\noindent
Let us set
\begin{eqnarray}
&&B(\xi,\eta\vert \txi,\teta)
=B(\xi)\ot B(\eta) \ot B(\tilde{\eta})^\ast\ot B(\tilde{\xi})^\ast,
\nonumber
\\
&&B_{\xi\eta,\tilde{\xi}\tilde{\eta}}
=\{b\in B(\xi,\eta\vert \txi,\teta)
\vert \ \et_i b=\ft_i b=0 \hbox{ for all $i$}\},
\nonumber
\\
&&B_{\xi\eta}^\la=\{b\in B(\xi)\ot B(\eta)
\vert \ \et_ib=0\hbox{ for all $i$ and $wt b=\la$}\},
\nonumber
\\
&&B^{\ast}_{\txi\teta}=\{b\in B(\teta)^{\ast}\ot B(\txi)^{\ast}
\vert \  \ft_ib=0\hbox{ for all $i$ and $wt b=-\la$}\}.
\nonumber
\end{eqnarray}
If we introduce the trivial action of $\et_i$ and $\ft_i$
on $B_{\xi\eta}^\la$ and $B^{\ast\la}_{\txi\teta}$, we have
the isomorphisms of affine crystals,
\begin{eqnarray}
&&
B(\xi)\ot B(\eta)\simeq \sqcup_{\la\in P^0_k}
B_{\xi\eta}^\la \ot B(\la),
\quad
B(\teta)^{\ast}\ot B(\txi)^{\ast}\simeq \sqcup_{\la\in P^0_k}
B(\la)^{\ast}\ot B_{\txi\teta}^{\ast\la},
\nonumber
\\
&&
B_{\xi\eta,\txi\teta}=
\sqcup_{\la\in P^0_k}
B_{\xi\eta}^\la \ot B_{\txi\teta}^{\ast\la}.
\nonumber
\end{eqnarray}
Then

\begin{prop}\cite{DJO}\label{singlet}
There is a bijection
\begin{eqnarray}
&&
B_{\xi\eta,\tilde{\xi}\tilde{\eta}}
\simeq
\{\hbox{$(\xi\eta,\tilde{\xi}\tilde{\eta})$ restricted paths}\},
\nonumber
\end{eqnarray}
given by
\begin{eqnarray}
&&
b_\xi\ot b\ot b_{-\tilde{\xi}}
\mapsto
(a(n))_{n\in\bz},
\nonumber
\\
&&
a(n-1)-a(n)=\hbox{wt} \ p(n)\hbox{ for any $n$},
\nonumber
\end{eqnarray}
where $b=(p(n))_{n\in\bz}\in B(\eta) \ot B(\tilde{\eta})^\ast$.
\end{prop}
\vskip3truemm

\noindent
In the proof of this proposition, we use the weight multiplicity
freeness of $B^{(N)}$.

Proposition \ref{singlet} motivates the following definition
of the space of states of the RSOS quantum spin chain,
\begin{eqnarray}
{\cal H}&=&\oplus{\cal H}_{\xi\eta,\tilde{\xi}\tilde{\eta}},
\nonumber
\\
{\cal H}_{\xi\eta,\tilde{\xi}\tilde{\eta}}
&=&
\{\hbox{$\uq$ singlet}\}\subset
V(\xi,\eta\vert \txi,\teta), \quad {\rm and}
\nonumber
\\
V(\xi,\eta \vert \txi,\teta)
&=&
V(\xi)\ot V(\eta)\ot V(\tilde{\eta})^{\ast a}\ot
V(\tilde{\xi})^{\ast a}.
\nonumber
\end{eqnarray}
Here the sum is over all
$(\xi,\eta),(\tilde{\xi},\tilde{\eta})\in P^0_l\times P^0_N$.
The tensor product is considered to be appropriately completed.
Then the crystallized space of states of the RSOS model
is $\sqcup B_{\xi\eta,\tilde{\xi}\tilde{\eta}}$,
the main object of study in this paper.

The creation operator is defined in terms of the spin $1/2$
vertex operators\cite{JMO},
\begin{eqnarray}
&&
\v_{\xi,\eta}^{\ast \xi',\eta'}(z)
=
\Phi^{\eta'}_{V^{(1)}\eta}(z)
\Phi_{\xi}^{\xi' V^{(1)}}(z)
\label{creation}
\end{eqnarray}
which act as
\begin{eqnarray}
V(\xi,\eta\vert \txi,\teta)
&\stackrel{\Phi_{\xi}^{\xi' V^{(1)}}(z)}{\ra}&
V(\xi')\ot V^{(1)}_z \ot V(\eta)\ot V(\tilde{\eta})^{\ast a}\ot
V(\tilde{\xi})^{\ast a}
\nonumber
\\
&\stackrel{\Phi^{\eta'}_{V^{(1)}\eta}(z)}{\ra}&
V(\xi',\eta' \vert \txi,\teta).
\nonumber
\end{eqnarray}
This operator obviously preserve the space ${\cal H}$,
since it is an intertwiner.

\section{Crystalline spinon basis for higher spin XXZ models}
\par
In this section we recall the results of \cite{NY}
in a slightly generalized form which will be needed for the case of the
RSOS models.
We denote by $\pek_{{\rm res},n}(r_1,r_2)$ the set of restricted paths
from $r_1$ to $r_2$ of length $n$ in the sense of \cite{NY}.
It is understood that $\pek_{{\rm res},0}=\{ \phi \}$.
We use the expression $B^{\rm XXZ}(p_n,\cdots,p_1)$ to represent
$B(p_n,\cdots,p_1)$ of section 2, \cite{NY}.
Let us set
\begin{eqnarray}
&&
B^{\rm XXZ}_{\geq m}(p_n,\cdots,p_1)
=\{
\v^{\ast p_n}_{j_n}\cdots\v^{\ast p_1}_{j_1}
\in B^{\rm XXZ}(p_n,\cdots,p_1)
\vert \ j_1\geq m\}.
\nonumber
\end{eqnarray}

The following theorem is proved in \cite{NY}.

\begin{theorem}\label{xxzbase}
Let $k$ be a positive integer and $0\leq l,r\leq k$.
Then there is an isomorphism of affine crystals
\begin{eqnarray}
&&
\sqcup_{n=0}^\infty\sqcup_{(p_n,\cdots,p_1)\in \pek_{{\rm res},n}(r,l)}
B^{\rm XXZ}(p_n,\cdots,p_1)
\simeq
B(\la_l)\ot B(\la_r)^\ast
\nonumber
\end{eqnarray}
given by
\begin{eqnarray}
1&\longmapsto&[[ \ ]]_r=b_{\la_r}\ot b_{-\la_r},
\nonumber
\\
\v^{\ast p_n}_{j_n}\cdots\v^{\ast p_1}_{j_1}
&\longmapsto&
[[j_n-p_n,\cdots,j_1-p_1]].
\nonumber
\end{eqnarray}
\end{theorem}

Although, in \cite{NY}, the statement of this theorem is only made for
the case $r=0$, the statement given above is actually proved.
In fact the bijectivity of the map is obvious, and the condition
$r=0$ is not used in the proof of the weight preservation.

\begin{cor}
The map in Theorem \ref{xxzbase} induces the bijection
preserving the affine weights:
\begin{eqnarray}
&&
\sqcup_{n=0}^\infty\sqcup_{(p_n,\cdots,p_1)\in \pek_{{\rm res},n}(r,l)}
B^{\rm XXZ}_{\geq p_1}(p_n,\cdots,p_1)
\simeq
B(\la_l),
\nonumber
\end{eqnarray}
where we make the identification:
\begin{eqnarray}
B(\la_l)&\simeq& B(\la_l)\ot b_{-\la_r}
\nonumber
\\
b&\longrightarrow& b\ot b_{-\la_r}.
\nonumber
\end{eqnarray}
\end{cor}
\par
\noindent

By a calculation similar to that in section 5 of \cite{NY}
we have

\begin{cor}
For any $0\leq r\leq k$, the character
${\rm ch}_j(q,z) \equiv {\rm tr}_{V(\la_{j})} (q^{-d} z^{h_1})$
is given by
\begin{eqnarray}
{\rm ch}_j(q,z)=\sum_{n=0}^\infty \sum_{m=0}^{n}
{1 \over (q)_{n-m} (q)_{m}}
\sum_{p} q^{h'(p)+mp_1} z^{n+r-2m},
\label{character}
\end{eqnarray}
where $p=(p_n,\cdots,p_{1})$ runs over all level
$k$ restricted fusion paths
from $r$ to $j$ and $h'(p)=\sum_{s=1}^{n-1}(n-s)H(p_{s+1},p_s)$.
\end{cor}

This corollary gives $k+1$ different expressions for the character
of $V(\la_j)$.

\section{Crystalline creation algebra}
\par
In this section we shall introduce the
algebra of creation operators of the RSOS model
at $q=0$. Unless otherwise stated we
use the notation of \cite{NY}.
For $p=0,1$ let us set $c(p)=1-p$.

\begin{definition}\label{cca}
The crystalline creation algebra $\ca^{\rm RSOS}$ is
the associative algebra with unity generated by
$\{\v^{\ast pp'}_{2j+c(p)} \ \vert
\ j\in\bz,\,p\in\{0,1\}\}\cup\{1\}$
over $\bz$ subject to the
following relations:
\begin{eqnarray}
&&
\v^{\ast p_2p_2^{\prime}}_{2j_2+c(p_2)}
\v^{\ast p_1p_1^{\prime}}_{2j_1+c(p_1)}+
\v^{\ast p_2p_2^{\prime}}_{2j_1+2s_{21}+c(p_2)}
\v^{\ast p_1p_1^{\prime}}_{2j_2-2s_{21}+c(p_1)}=0,
\label{crels}
\end{eqnarray}
where
\begin{eqnarray}
s_{21}=-1+H(p_2,p_1)+H(p_2^{\prime},p_1^{\prime}).
\nonumber
\end{eqnarray}
\end{definition}
\vskip3truemm

The algebra $\ca^{\rm RSOS}$ is naturally graded by
\begin{eqnarray}
&&
\ca^{\rm RSOS}=\oplus_{n=0}^\infty \ca_n^{\rm RSOS}
\nonumber
\\
&&
\ca^{\rm RSOS}_n=\sum
\bz
\v_{2j_n+c(p_n)}^{\ast p_np_n^{\prime}}
\cdots
\v_{2j_1+c(p_1)}^{\ast p_1p_1^{\prime}},
\quad
\ca^{\rm RSOS}_0=\bz.
\nonumber
\end{eqnarray}

In the following, we denote the crystalline
creation algebra of the spin $l/2$ XXZ model\cite{NY}
by $\ca^{\rm XXZ}$,
where the integer $l=k-N$ is associated with the RSOS model
as in section 1.
For $p=(p_n,\cdots,p_1)$ and
$p^\prime=(p_n^\prime,\cdots,p_1^\prime)$ in $\{0,1\}^n$,
let us set
\begin{eqnarray}
&&B^{\rm RSOS}(p \vert p^\prime)
\nonumber \\
&&=\{
\v_{2j_n+c(p_n)}^{\ast p_np_n^{\prime}}
\cdots
\v_{2j_1+c(p_1)}^{\ast p_1p_1^{\prime}}
\vert
\hbox{ $(j_n,\cdots,j_1)$ satisfies the condition (\ref{nor})}\},
\nonumber
\\
&&B^{\rm RSOS}_{\geq0}(p \vert p^\prime)
\nonumber \\
&&=\{
\v_{2j_n+c(p_n)}^{\ast p_np_n^{\prime}}
\cdots
\v_{2j_1+c(p_1)}^{\ast p_1p_1^{\prime}}
\in B^{\rm RSOS}(p \vert p^\prime)
\vert
j_1\geq H(p_1^{\prime},c(p_1))
\}.
\nonumber
\end{eqnarray}
If $n=0$, we define $(p_n, \cdots, p_1)=\phi$ and
$B^{\rm RSOS}(\phi\vert \phi)=\{ 1 \}$.
The condition is
\begin{eqnarray}
&&
j_n-I_n-I_n^{\prime}\geq\cdots\geq
j_2-I_2-I_2^{\prime}\geq j_1,
\label{nor}
\end{eqnarray}
where
\begin{eqnarray}
I_m
&=&
I_m(p_m,\cdots,p_1)=\sum_{s=1}^{m-1}H(p_{s+1},p_s),
\nonumber
\\
I_m^{\prime}
&=&
I_m(p_m^{\prime},\cdots,p_1^{\prime})=
\sum_{s=1}^{m-1}H(p_{s+1}^{\prime},p_s^{\prime}).
\nonumber
\end{eqnarray}

\noindent
If we set
\begin{eqnarray}
&&
\psi_A(j)=\v^{\ast pp'}_{2j+c(p)},
\quad
A=(p,p')
\nonumber
\end{eqnarray}
and $A_i=(p_i,p_i^{\prime})$ $(i=0,1)$,
$s_{A_2A_1}=s_{21}$, then the commutation relations
in Definition 3 can be written as
\begin{eqnarray}
&&
\psi_{A_2}(j_2)
\psi_{A_1}(j_1)
+
\psi_{A_2}(j_1+s_{A_2A_1})
\psi_{A_2}(j_2-s_{A_2A_1})
=0.
\nonumber
\end{eqnarray}

\noindent
Then the condition (\ref{nor}) is nothing but the normality
condition in the sense of Definition 2 in \cite{NY}.
Hence by Corollary 2 of \cite{NY} we have

\begin{theorem}\label{linear}
$\sqcup_{p,p'\in\{0,1\}^n}B^{\rm RSOS}(p \vert p^\prime)$
is a $\bz$ linear base of $\ca^{\rm RSOS}_n$.
\end{theorem}


\begin{definition}
Let us define the weight of an element of
$B^{\rm RSOS}(p \vert p^\prime)$ by
\begin{eqnarray}
wt(\v_{2j_n+c(p_n)}^{\ast p_np_n^{\prime}}
\cdots
\v_{2j_1+c(p_1)}^{\ast p_1p_1^{\prime}})&=&
\sum_{s=1}^nwt(\v_{2j_s+c(p_s)}^{\ast p_sp_s^{\prime}})
\nonumber
\\
wt(\v_{2j+c(p)}^{\ast pp^{\prime}})&=&
-j\delta,
\nonumber
\\
wt1&=&0.
\nonumber
\end{eqnarray}
\end{definition}

\noindent
We introduce the structure of a crystal in
$B^{\rm RSOS}(p \vert p^\prime)$
such that $\et_i$, $\ft_i$ $(i=0,1)$ act as $0$ on any element,
and $B^{\rm RSOS}(p \vert p^\prime)$ has the weights defined above.

Note that the commutation relations (\ref{crels}) are the same as those
of
$\v^{\ast p_2^{\prime}}_{2j_2+c(p_2)}$ and
$\v^{\ast p_1^{\prime}}_{2j_1+c(p_1)}$.
Hence, by Theorem 1 in \cite{NY} and Theorem \ref{linear} above, we have

\begin{theorem}\label{embed}
There is an embedding of the algebra
$\ca^{\rm RSOS}\ra \ca^{\rm XXZ}$ given by
\begin{eqnarray}
&&
\v_{2j+c(p)}^{\ast pp^{\prime}}
\mapsto
\v_{2j+c(p)}^{\ast p^{\prime}}.
\nonumber
\end{eqnarray}
Under this embedding
$B^{\rm RSOS}(p \vert p^\prime)$
is mapped into
$B^{\rm XXZ}(p^\prime)$ for $p,p^\prime\in\{0,1\}^n$.
Moreover, the weight in $d$ is preserved under this embedding.
\end{theorem}

\section{Crystalline spinon basis for the RSOS models}
\par
In this section we give a new parametrization of a base of
the set of highest weight vectors in the tensor products of
two integrable highest weight $\uq$ modules in terms
of crystalline spinons.
Let us consider the integers $k,N$ and $l$ as introduced in section 1.
We now state the main theorem of this paper.

\begin{theorem}\label{base}
For $r_1,l_1\in\{0,\cdots,l\}$
and
$r_2,l_2\in\{0,\cdots,N\}$,
there is an isomorphism of affine crystals
\begin{eqnarray}
&&
\sqcup_{n=0}^\infty
\sqcup_{p\in \pel_{{\rm res},n}(r_1,l_1)}
\sqcup_{p^\prime\in \peN_{{\rm res},n}(r_2,l_2)}
B^{\rm RSOS}(p \vert p^\prime)
\simeq
B_{\la_{l_1}^{(l)}\la_{l_2}^{(N)},\la_{r_1}^{(l)}\la_{r_2}^{(N)}}
\nonumber
\end{eqnarray}
given by
\begin{eqnarray}
1&\longmapsto&[[ \ ]]_{r_1,r_2}=
b_{\la_{r_1}^{(l)}}\ot b_{\la_{r_2}^{(N)}}\ot
b_{-\la_{r_2}^{(N)}}\ot b_{-\la_{r_1}^{(l)}},
\nonumber
\\
\v_{2j_n+c(p_n)}^{\ast p_np_n^{\prime}}
\cdots
\v_{2j_1+c(p_1)}^{\ast p_1p_1^{\prime}}
&\longmapsto&
b_{\la_{l_1}^{(l)}}\ot
\v_{2j_n+c(p_n)}^{\ast p_n^{\prime}}
\cdots
\v_{2j_1+c(p_1)}^{\ast p_1^{\prime}}
\ot b_{-\la_{r_1}^{(l)}}.
\nonumber
\end{eqnarray}
\end{theorem}
\vskip4truemm

\noindent
By restricting the above map to the
set of elements of the form
$b\ot b^\prime\ot b_{-\la_{r_2}^{(N)}}\ot b_{-\la_{r_1}^{(l)}}$ in
$B_{\la_{l_1}^{(l)}\la_{l_2}^{(N)},\la_{r_1}^{(l)}\la_{r_2}^{(N)}}$, we have

\begin{cor}\label{hbase}
The map in Theorem \ref{base} induces the bijection
preserving the affine weights:
\begin{eqnarray}
&&
\sqcup_{n=0}^\infty
\sqcup_{p\in \pel_{{\rm res},n}(r_1,l_1)}
\sqcup_{p^\prime\in \peN_{{\rm res},n}(r_2,l_2)}
B^{\rm RSOS}_{\geq0}(p \vert p^\prime)
\simeq
B^{\la_{r}^{(k)}}_{\la_{l_1}^{(l)}\la_{l_2}^{(N)}},
\nonumber
\end{eqnarray}
where $r=r_1+r_2$, and we make the identification:
\begin{eqnarray}
B^{\la_{r}^{(k)}}_{\la_{l_1}^{(l)}\la_{l_2}^{(N)}}&\simeq&
B^{\la_{r}^{(k)}}_{\la_{l_1}^{(l)}\la_{l_2}^{(N)}}\ot
b_{-\la_{r_2}^{(N)}}\ot b_{-\la_{r_1}^{(l)}},
\nonumber
\\
b&\longrightarrow& b\ot b_{-\la_{r_2}^{(N)}}\ot b_{-\la_{r_1}^{(l)}}.
\nonumber
\end{eqnarray}
\end{cor}
\par

Let us define
\begin{eqnarray}
&&
V_{l_1,l_2}^{r}=
\{
v\in V(\la_{l_1}^{(l)})\ot V(\la_{l_2}^{(N)})
\vert
e_i v=0 \ (i=0,1), \ \hbox{wt}v=\la_r^{(k)}
\}.
\nonumber
\end{eqnarray}
By calculations similar to those in
section 5 of \cite{NY} we have

\begin{cor}
Assume the same conditions as those in Theorem \ref{base}.
Then we have
\begin{eqnarray}
&&
\hbox{tr}_{V_{l_1,l_2}^{r}}(q^{-d})
=
\sum_{n=0}^\infty
{1\over (q)_n}
\sum_{p\in \pel_{{\rm res},n}(r_1,l_1)}
\sum_{p'\in \peN_{{\rm res},n}(r_2,l_2)}
q^{h'(p)+h'(p')+nH(p_1',c(p_1))} .
\nonumber
\end{eqnarray}
In particular, if $r_1r_2=0$, then
\begin{eqnarray}
&&
\hbox{tr}_{V_{l_1,l_2}^{r}}(q^{-d})=
\sum_{n=0}^\infty{1\over (q)_n}
K^{l}_{r_1,l_1}(n)K^{N}_{r_2,l_2}(n),
\nonumber
\end{eqnarray}
where the polynomial $K^k_{r,j}(n)$ is that defined in (\ref{poly})
of section 5.
\end{cor}
\vskip5truemm

We remark that the branching coefficient\cite{KW}
is obtained from
$\hbox{tr}_{V_{l_1,l_2}^{r}}(q^{-d})$
as
$q^{
s_{l_1}^{(l)}+
s_{l_2}^{(N)}-
s_{r}^{(k)} }
\hbox{tr}_{V_{l_1,l_2}^{r}}(q^{-d})$, where
$s^{(l)}_m={(m+1)^2\over 4(l+2)}-{1\over8}$.
\vskip3truemm

{\bf Example.}
For the simplest case, $l=N=1$, one has
\begin{eqnarray}
&&\hbox{tr}_{V_{0,0}^{0}}(q^{-d})=
\sum_{n=0}^{\infty} {q^{2 n^2} \over (q)_{2n}},
\nonumber \\
&&\hbox{tr}_{V_{1,1}^{0}}(q^{-d})=
\sum_{n=0}^{\infty} {q^{2n(n+1)} \over (q)_{2n+1}},
\nonumber \\
&&\hbox{tr}_{V_{0,1}^{1}}(q^{-d})=
\sum_{n=0}^{\infty} {q^{n(2n-1)} \over (q)_{2n}}=
\sum_{n=0}^{\infty} {q^{n(2n+1)} \over (q)_{2n+1}}.
\nonumber
\end{eqnarray}
These are well known fermionic character formulas
for the Ising model with $h=0,1/2$ and $1/16$ up to the
normalization $q^{h-1/48}$ \cite{KMM}.
\vskip3truemm

\noindent
Theorem \ref{base} follows from
Theorem \ref{embed} and the following lemma.

\begin{lemma}\label{singlecity}
Let
$
b=
b_{\la_{l_1}^{(l)}}\ot
\v_{2j_n+c(\ep_n)}^{\ast p_n^{\prime}}
\cdots
\v_{2j_1+c(\ep_1)}^{\ast p_1^{\prime}}
\ot b_{-\la_{r_1}^{(l)}}
$
be an element of
$B(\la_{l_1}^{(l)},\la_{l_2}^{(N)}\vert \la_{-r_1}^{(l)},\la_{-r_2}^{(N)})$.
Then $\tilde{x}_ib=0$ for $i=0,1$ and $x=e,f$ if and only if the path
$(\ep_n,\cdots,\ep_1)$ beginning at $r_1$
is an element of $\peN_{{\rm res},n}(r_1,l_1)$.
\end{lemma}

\noindent
The following statement is easily proved.

\begin{lemma}\label{sub}
Let $b'=b^{(1)}_{c(\mu_n)}\ot\cdots \ot b^{(1)}_{c(\mu_1)}$
be an element of $B^{(1)\ot n}$. Then
\begin{description}
\item[(i)] If $\et_1b'=\ft_1b'=0$, the path $(\mu_n,\cdots,\mu_1)$
beginning at $0$ never falls below $0$.
\item[(ii)] If $\et_0b'=\ft_0b'=0$, the path $(\mu_n,\cdots,\mu_1)$
beginning at $N$ never rises above $N$.
\end{description}
\end{lemma}
\vskip5truemm

\noindent
Proof of Lemma \ref{singlecity}.
Note that the condition $\et_ib=\ft_ib=0$ for $i=0,1$
is equivalent to
\begin{eqnarray}
&&
\tilde{x}_1(
b^{(1)\ot l_1}_0\ot
b^{(1)}_{c(\ep_n)}\ot \cdots \ot b^{(1)}_{c(\ep_1)}
\ot b^{(1)\ot r_1}_1)=0
\hbox{ for $x=e,f$},
\nonumber
\\
&&
\tilde{x}_0(
b^{(1)\ot N-l_1}_1\ot
b^{(1)}_{c(\ep_n)}\ot \cdots \ot b^{(1)}_{c(\ep_1)}
\ot b^{(1)\ot N-r_1}_0)=0
\hbox{ for $x=e,f$}.
\nonumber
\end{eqnarray}
Then Lemma \ref{singlecity} follows from Lemma \ref{sub}.
$\Box$.

\section{Some other applications}
To this point we have shown that the crystalline spinon picture
can be naturally extended to the case of RSOS models.
In this section, we discuss some additional applications of our
construction.

First, we discuss various formulas for the string function
of integrable highest weight ${\widehat {sl_2}}$ modules
which follow from the spinon character formulas of \cite{NY}.
Then, considering the limiting forms of these formulas, one can obtain
the $sl_2$ parafermion characters for $c=3k/(k+2)-1$
as suitable large $n$ limits of the polynomials $K^k_{r,j}(n)$.
Finally, we comment on another large $n$ limit of the polynomials
$K^k_{r,j}(n)$ and their relation to the Virasoro minimal model
characters with $c=1-6/(k+1)(k+2)$.

Let us define $K^k_{r,j}(n)$, ($0 \leq r,j \leq k$).
\begin{eqnarray}
&&
K^{k}_{r,j}(n)=\sum_{p \in {\cal P}^{(k)}_{{\rm res},n}(r,j)} q^{h'(p)},
\label{poly}
\end{eqnarray}
where $h'(p)$ is given by
$$
h'(p)=\sum_{i=1}^{n-1} (n-i) H(p_{i+1},p_{i}).
$$
By definition, $K^k_{r,j}(n)$ is a polynomial in $q$
with non-negative integer coefficients.

One can prove the following formula
\begin{eqnarray}
&&K^k_{r,j}(n)=G^{k+2}_{r,j+1}(n)-G^{k+2}_{r,-j-1}(n)
\nonumber \\
&&G^l_{r,s}(n)=\sum_{m \in \bz}
q^{m s+m^2 l}
\left[
\begin{array}{c}
n \\
{s+2ml+n-r-1 \over 2}
\end{array} \right].
\nonumber
\end{eqnarray}

The string function\footnote{We normalize it as $c^{\la}_{\mu}(q)=1+{\cal
O}(q)$.}
$c^{j}_{\mu}(q)$ is the character of
the weight $h_1=\mu$ subspace of the integrable module $V(\la_j)$.
It is easy to read the string functions from
the spinon character formula\cite{BLS2,NY},
$$
c^{j}_{\mu}(q)=
q^a \sum_{n=\vert \mu \vert}^{\infty} {K^k_{0,j}(n) \over
(q)_{n-\mu \over 2}(q)_{n+\mu \over 2}},
$$
where $q^a$ is a suitable normalization factor depending on $\mu$.

It should be noted that in these formulas the symmetry relations
$$
c^j_\mu=c^j_{-\mu}=c^j_{\mu+2 k \bz}=c^{k-j}_{k-\mu}
$$
are not manifest.
Considering this fault as a virtue, one can derive infinitely
many $q$-series identities.
For instance, at level $k=1$, all string functions are equal, and
one obtains
$$
\sum_{n=0}^{\infty} {q^{n(n+m)} \over (q)_n (q)_{n+m}}={1 \over (q)_{\infty}},
$$
for any $m \geq 0$. The expression on the rhs may be identified with
the $m \rightarrow \infty$ limit of the lhs.
Similarly, for $k=2$, one has
\begin{eqnarray}
&&(q)_{\infty} c^0_0=
\lim_{n \rightarrow \infty} q^{-2 n^2} K^2_{0,0}(4 n),
\nonumber \\
&&(q)_{\infty} c^0_2=
\lim_{n \rightarrow \infty} q^{-(2 n^2+2n+1)} K^2_{0,0}(4 n+2),
\nonumber \\
&&(q)_{\infty} c^1_1=
\lim_{n \rightarrow \infty} q^{-(2 n^2+n)} K^2_{0,1}(4 n+1).
\nonumber
\end{eqnarray}
These are essentially the Virasoro characters for $c=1/2$ and $h=0,1/2$
and $1/16$.

In general, the large $n$ behavior of $K^k_{r,j}(n)$ is described by

\begin{lemma}
For $(a)$ $0 \leq i < k-j$ and $(b)$ $k-j \leq i < k$, put
\begin{eqnarray}
&(a)& K^k_{0,j}(2nk+2i+j)= q^{k n^2+j n+i(2n+1)} {\bar K}^k_{0,j}(2nk+2i+j),
\nonumber \\
&(b)& K^k_{0,j}(2nk+2i+j)= q^{k n^2+j n+i(2n+2)+j-k} {\bar
K}^k_{0,j}(2nk+2i+j).
\nonumber
\end{eqnarray}
Then, ${\bar K}^k_{0,j}(2nk+2i+j)=1+{\cal O}(q)$, and has a limit as
$n \rightarrow \infty$ in the form of formal power series in $q$.
\end{lemma}

Proof.
The lowest degree term comes from the following path
\begin{eqnarray}
&(0^k 1^k)^n 0^{i+j} 1^i,
&(0 \leq i < k-j)
\nonumber \\
&(0^k 1^k)^n 0^k 1^i 0^{i+j-k},
&(k-j \leq i < k)
\nonumber
\end{eqnarray}
and the number of paths of fixed degree is finite and independent
of n (for $ n >>0 $), since those paths can differ from the lowest
degree path only near end points.$\Box$

By this lemma, the limiting form of the infinite sequence
$$
c^j_\mu=c^j_{\mu+2k}=c^j_{\mu+4k}=\cdots,
$$
takes the form
$$
c^j_\mu=
{1 \over (q)_{\infty} }
\lim_{n \rightarrow \infty} {\bar K}^k_{0,j}(2kn+\mu) .
$$
In particular $\lim_{n \rightarrow \infty} {\bar K}^k_{0,j}(2kn+\mu)$
is an analytical function on $\vert q \vert<1$.
Since $(q)_{\infty} c^j_\mu$ is the parafermion character
\footnote{This is also normalized as ${\rm ch}^j_{\mu}=1+{\cal O}(q)$.}
${\rm ch}^j_\mu$ with $c=3k/(k+2)-1$, we have

\begin{prop}
The large $n$ limit of the
lower degree terms in $K^k_{0,j}(n)$
gives the parafermion character
$$
{\rm ch}^j_\mu=(q)_{\infty} c^j_\mu=
\lim_{n \rightarrow \infty} {\bar K}^k_{0,j}(2kn+\mu).
$$
\end{prop}

The sum in $K^k_{r,j}(n)$ looks like the usual path realization
of Virasoro minimal model characters \cite{ABF}. Hence
it is natural to seek the relation between these two.
In doing this, let us first introduce the 1D configuration
sum of the ABF model in regime II or III by
\begin{eqnarray}
&&
X^k_{j,r}(n,q)=\sum_{p\in \pek_{res,n}(j,r)}
q^{\omega'(c(p))},
\nonumber
\end{eqnarray}
where
\begin{eqnarray}
&&
\omega'(p)=\sum_{i=1}^{n-1} i \tilde{H}(p_{i+1},p_i).
\nonumber
\end{eqnarray}
Here $\tilde{H}(0,1)=-1$, $\tilde{H}(0,0)=\tilde{H}(1,1)=\tilde{H}(1,0)=0$,
and $c(p)=(c(p_i))$.
Let us next set
\begin{eqnarray}
&&
b^j_{r,s} (q)={\rm tr}_{V^j_{r, s}} (q^{-d}),
\nonumber
\end{eqnarray}
which is the branching coefficient up to the power of $q$, as mentioned in
section 4 for
$$
V(\la_r^{(k-1)})\ot V(\la_{s}^{(1)})\simeq
\oplus_{j}V_{r,s}^{j}\ot V(\la_{j}^{(k)}),
$$
where $j \equiv r+s$ (mod. 2).
We write $K^k_{j,r}(n,q)=K^k_{j,r}(n)$.
Then we have

\begin{prop}
\begin{description}
\item[]
\item[(i)]
$K^k_{j,r}(n,q)=X^k_{r,j}(n,q^{-1})$.
\item[(ii)]
For $\vert q \vert<1$ we have
\begin{eqnarray}
&&
b^j_{r,s} (q)=
\lim_{n\ra\infty}q^{n(n-r)}K^k_{r+s,j}(2n,q^{-1}).
\nonumber
\end{eqnarray}
\end{description}
\end{prop}
\par
\noindent
Proof.
The proof of (i) follows immediately from the definitions.
Let us prove (ii). Using the formulation of \cite{DJO} we have
\begin{eqnarray}
&&
b^j_{r, s}(q)=
\lim_{n \rightarrow \infty}
\sum_{p \in {\cal P}^{(k)}_{{\rm res},2n}(j,r+s)}
q^{{\omega'}(c(p))-{\omega'}(p_{gr}^r)}=
\lim_{n \rightarrow \infty}
q^{-{\omega'}(p_{gr}^r)}
X^k_{j,r+s}(2n,q),
\nonumber
\end{eqnarray}
where $p_{gr}^r(n)=\ep(n+r)$ and $\ep(n)={1\over2}(1-(-1)^n)$.
Since $\omega'(p_{gr}^r)=-n(n-r)$,
(ii) follows from (i).
$\Box$
\vskip3truemm

\noindent
It follows from (i) of this proposition that
$\lim_{n \rightarrow \infty}{\bar K}^k_{0,j}(2kn+\mu)$ discussed
above is the 1D configuration sum in the thermodynamic limit
of the ABF model in regime II.

\section{Discussion}
In this paper we introduced the crystalline creation
algebra for RSOS models. Using this we have given a description
of the set of highest weight elements in the tensor product
of crystals of integrable highest weight $\uq$ modules
in terms of crystalline spinons. This crystalline spinon basis
leads us to the fermionic type formulas of the branching functions.
We have found that the infinite system of string functions obtained from
the spinon basis converge to the one dimensional configuration sums
(divided by $(q)_\infty$) of the ABF model in regime II.
This allows a path description of
the parafermion characters.

The fermionic formulas for the branching coefficients associated with
the ABF model were first proved in \cite{Berk}.
In that work, however,
the relation of the character formula with the underlying quasi-particle
structure is rather implicit or there was no description given
of the path counting with weights.
Recently O. Warnaar gave another proof by counting weighted paths
based on the Fermi gas picture. He also discussed its relation
with the Bethe Ansatz solutions \cite{Wa}.
Our results are the rigorous generalization of the character
formulas presented in \cite{Berk,Wa}
to the case of general RSOS models, directly related with
the quasi-particle structure of the model.

The relation between the one dimensional configuration sums of
the ABF model in regime II and the parafermion characters found here
is consistent with the results of \cite{BR}. It is an interesting problem
to determine whether this relation can be generalized to the case of general
RSOS models and whether there are corresponding spinon type
series.

\appendix
\section{Appendix}
Here we briefly explain the derivation of the commutation relations
of the creation operators at $q=0$ in section 3.
In the following calculations, we assume that the creation
operators have a well defined $q\ra0$ limit. We remark that
in the main text we do not use this assumption.

We use the notation and results of \cite{JMO}.
The commutation relations of creation operators,
as determined in \cite{JMO}, are:
\begin{eqnarray}
&&
\v^{\ast \xi''\eta''}_{\xi'\eta'}(z_1)
\v^{\ast \xi'\eta'}_{\xi\eta}(z_2)
\nonumber
\\
&&
=\sum_{\zeta,\kappa}
\v^{\ast \xi''\eta''}_{\zeta\kappa}(z_2)
\v^{\ast \zeta\kappa}_{\xi\eta}(z_1)
W^1_l(
\begin{array}{cc} \xi & \xi' \\ \zeta & \xi'' \end{array}
\vert {z_1\over z_2})
W^{\ast 1}_N(
\begin{array}{cc} \eta & \eta' \\ \kappa & \eta'' \end{array}
\vert {z_1\over z_2}).
\nonumber
\end{eqnarray}

As in the case of higher spin XXZ (cf. Appendix A in \cite{NY}),
the $q\ra0$ limit of the Boltzmann weights does not depend
on the initial weight but on the difference
of weights if we remove the fractional powers.
So let us set
\begin{eqnarray}
&&
\v^{\ast pp'}(z)=
z^{\Delta_\xi-\Delta_{\xi'}+\Delta_\eta-\Delta_{\eta'}}
\v^{\ast \xi'\eta'}_{\xi\eta}(z)\vert_{q=0},
\nonumber
\end{eqnarray}
where $p,p'=0,1$ are defined by
\begin{eqnarray}
&&
\xi'-\xi=(-1)^p(\La_1-\La_0),
\quad
\eta'-\eta=(-1)^{p'}(\La_1-\La_0).
\nonumber
\end{eqnarray}
Then taking the $q\ra0$ limit of the commutation relations,
we have
\begin{eqnarray}
&&
\v^{\ast p_1p_1'}(z_1)\v^{\ast p_2p_2'}(z_2)
=-
\Big(
{z_1\over z_2}
\Big)^{-1+H(p_1,p_2)+H(p_1',p_2')}
\v^{\ast p_1p_1'}(z_2)\v^{\ast p_2p_2'}(z_1),
\nonumber
\end{eqnarray}
where the function $H$ is defined by
\begin{eqnarray}
&&
H(1,0)=1,
\quad
H(i,j)=0 \hbox{ otherwise}.
\nonumber
\end{eqnarray}

The algebra in Definition \ref{cca} can be obtained by
the following mode expansion.
\begin{eqnarray}
&&\v^{\ast pp'}(z)=\sum_{j \in \bz} \v^{\ast pp'}_{2j+c(p)} z^j.
\nonumber
\end{eqnarray}
The reason that the mode expansion takes this form
is explained as follows.
The creation operator is expressed as the composition
of type I and type II vertex operators given in (\ref{creation}).
Here the type I vertex operator preserves the crystal lattice, and
the type II vertex operator is the same as the creation
operators of the spin $l/2$ XXZ model.
Noting that a highest weight vector in $V(\xi)\ot V(\eta)$ is dominated
in the $q\ra0$ limit by the element of the form
$u_\xi\ot v$ $(v\in V(\eta))$, we expect a mode expansion as above.


\clearpage
\begin{thebibliography}{99}
\bibitem{ABF}Andrews, G., Baxter, R. and Forrester, P.,
{Eight-vertex SOS model and generalized Rogers-Ramanujan
type identities,
J. Stat. Phys.{\bf 35}, 193-266 (1984).}
\bibitem{BR} Bazhanov, V.V. and Reshetikhin, N. Yu.,
{Scattering amplitudes in offcritical models and RSOS integrable models,
Progress of Theor. Phys. Supple. {\bf 102}, 301-318 (1990).}
\bibitem{Berk} Berkovich, A.,
{Fermionic counting of RSOS-states and Virasoro character formulas
for the unitary minimal series $M(\nu,\nu+1)$. Exact results,
hep-th/9403073 (1994).}
\bibitem{BPS} Bernard, D., Pasquier, V., Serban, D.
{Spinons in conformal field theory,
Nucl. Phys. B{\bf 428}, 612-628 (1994).}
\bibitem{BLS1} Bouwknegt, P., Ludwig, A., Schoutens, K.
{Spinon bases, Yangian symmetry and fermionic representations
of Virasoro characters in conformal field theory,
Phys. Lett. B {\bf 338}, 448-456 (1994).}
\bibitem{BLS2} Bouwknegt,P., Ludwig, A., Schoutens, K.
{Spinon bases for higher level SU(2) WZW models,
preprint, hep-th/9412108 (1994).}
\bibitem{DFJMN} Davies, B., Foda, O., Jimbo, M., Miwa, T.,
Nakayashiki, A.
{Diagonalization of the XXZ hamiltonian by vertex operators,
CMP {\bf 151}, 89-153 (1993).}
\bibitem{DJKMO}Date, E., Jimbo, M., Kuniba, A., Miwa, T., Okado, M.,
{Exactly solvable SOS models II, proof of the star-triangle
relation and combinatorial identities, Adv. Stud. in Pure Math.
{\bf 16}, 17-122 (1988).}
\bibitem{DJO}Date, E., Jimbo, M., Okado, M.,
{Crystal base and q vertex operators, CMP {\bf 155}, 47-69 (1993).}
\bibitem{FQ1} Foda, O. and Quano, Y-H.,
{Polynomial identities of the Rogers-Ramanujan type,
hep-th/9407191}
\bibitem{FQ2} Foda, O. and Quano, Y-H.,
{Virasoro character identities from the Andrews-Bailey construction,
hep-th/9408086}
\bibitem{IIJMNT}Idzumi, M., Iohara, K., Jimbo, M., Miwa, T.,
Nakashima, T., Tokihiro, T.,
{Quantum affine symmetry in vertex models, IJMPA {\bf 8},
1479-1511 (1993).}
\bibitem{JMO}Jimbo, M., Miwa, T., Ohta, Y.,
{Structure of the space of states in RSOS models,
IJMPA {\bf 8},
1457-1477 (1993).}
\bibitem{KW}Kac, V., Wakimoto, M.,
{Modular and conformal invariance constraints in representation
theory of affine algebras, Adv. in Math. {\bf 70},
156-236 (1988).}
\bibitem{K1}Kashiwara, M.,
{On crystal bases of the q-analogue of universal
enveloping algebras, Duke Math. J. {\bf 63} No.2, 465-516 (1991).}
\bibitem{KKMMNN1}Kang, S-J., Kashiwara, M., Misra, K.,
Miwa, T., Nakashima, T. and  Nakayashiki, A.
{Affine crystals and vertex models,
Int. J. Mod. Phys. A {\bf 7}, Suppl. 1A, 449-484 (1992).}
\bibitem{KMM}Kedem, R., McCoy, B.M. and Melzer, E.,
{The sums of Rogers, Schur and Ramanujan and the Bose-Fermi
correspondence in $1+1$-dimensional quantum field theory,
hep-th/9304056  (1993).}
\bibitem{K}Kirillov, A. N.,
{Dilogarithm identities, preprint,hep-th/9408113  (1994).}
\bibitem{KNS} Kuniba, A., Nakanishi, T. and Suzuki, J.,
{Characters in conformal field theories from thermodynamic
Bethe Ansatz,
Mod. Phys. Lett. A {\bf 8}  (1993).}
\bibitem{NY} Nakayashiki, A. and Yamada, Y.,
{Crystallizing the spinon basis,
Kyushu Univ. preprint (1995), hep-th/9504052}
\bibitem{Wa} Warnaar, S.O.,
{Fermionic solutions of the Andrews-Baxter-Forrester model I:
unification of TBA and CTM methods, hep-th/9501134}
\end{thebibliography}
\end{document}